# Repeat-pass SAR Interferometry Experiments with Gaofen-3: A Case Study of Ningbo Area

Tao Zhang, Xiaolei Lv, Bing Han, Bin Lei and Jun Hong

Key Laboratory of Technology in Geo-spatial Information Processing and Application System, Institute of Electronics, Chinese Academy of Sciences, Beijing 100190, China

E-mail: forzhangtao@163.com

**Abstract**

This paper reports the repeat-pass interferometric SAR results of Gaofen-3, a Chinese civil SAR satellite, acquired in November 2016 and March 2017 from Ningbo area. With the spatial baseline about 600 m and time baseline 116 days, the coherence of the two images still achieve good enough to generate the digital elevation model (DEM). During the InSAR processing, we compared several baseline estimating methods and obtained a good flat-earth phase removed interferogram map. By using the latest SAR interferogram filter and phase unwrapping method we proposed, we improved the coherence up to 0.8 in urban area and obtained a high quality DEM in Ningbo area. In addition, we evaluated the elevation model by comparing with the elevation values extracted from SRTM. And the result shows that accuracy of the elevation map is about 5 m (RMS) in plane area and is about 22m (RMS) in mountainous region, which demonstrated that Gaofen-3 has the powerful ability of repeat-pass SAR Interferometry.

## 1. Introduction

China has launched several SAR satellites in the last decade. However, most of them are aimed at agricultural, forestry and urban planning or target observing. As a first Chinese C-band multi-polarization high-resolution SAR satellite, Gaofen-3 (GF-3) , launched on August 10, 2016 in Taiyuan satellite center, is designed to provide high resolution imaging of remote sensing data in order to monitor the global sea and land information throughout the day and weather. The orbit is repeated with 29 days and the SAR has 12 kinds of imaging mode, resolution from 1 to 500 m, and can get the imaging width from 10 to 650 km, with functions of detailed and roughly observation. In the end of January 2017, the satellite was formally put into use. And up to March 25, 2017, it has obtained nearly 100 thousand scenes of multi-polarization SAR imaging in ocean and land to provide data support for the domestic and foreign resources census, the typhoon early warning, disaster evaluation, crop yield estimation, polar exploration, and many other applications. Due to vast and diverse terrain in China, terrain mapping faces many challenges. And geological disasters like landslide and debris flow that in relation to terrain deformation occurred occasionally. Moreover, overexploitation of the groundwater, urban subway construction may cause the subsidence of terrain. Therefore, the obtaining of the wide range high-precision ground elevation and deformation information plays an important role in safety production and disaster warning.

InSAR technique has proven reliable for elevation mapping and deformation monitoring [1,2], and has been used widely over the world. Benefiting from the high precision of the satellite attitude, orbit control and the system calibration, the acquired image pair of GF-3 has high coherence and has the possibility of InSAR processing. In the case this paper studied, the GF-3 satellite SAR imaging quality is good and stable, and even with the 600 m special baseline and the four-month time baseline, its coherence is up to 0.8 in some urban area.

In this paper, we use the repeat-pass imaging data of Ningbo area to carry out the InSAR processing. In the following, we focus on the detailed analysis of baseline estimation, coherence enhancement and phase unwrapping for the data. In addition, the inversion of elevation is evaluated with reference to the elevation values extracted from the SRTM. The experiment results show that the mean square error of the elevation is round 5m.

**2.  Description of the studied area and the data pre-processing**

Ningbo ranges in latitude from 28°51' to 30°33' N and in longitude from 120°55' to 122°16' E, is a sub-provincial city in northeast Zhejiang province in China. In the SAR image, the dark areas are water. The north-south mountain is located in the southeast of the imaging area, and white area in the image center is urban area, and the west of the image is also the mountain.

As shown in Figure 1, we applied the interferometric processes to the two descending-orbit GF-3 satellite images at HH polarization, acquired in November 13, 2016 and March 9, 2017, in Fine StripMap mode. The resolution of the SAR images is about 2.5 m in range direction and 2 m in azimuth direction, and the images both are 10000 × 10000 pixels in size. We set the early one as the master image and the later one as the slave image. Owing to the stable of spaceborne SAR, the image pair only has an offset of 200 pixels in range direction, and we co-registered the image pair by using intensity correlation based polynomial method. Then after the resampling the slave image according to the offset polynomial, the interferogram map was obtained by the conjugate multiplication of the master and slave image with 4 by 4 looks.

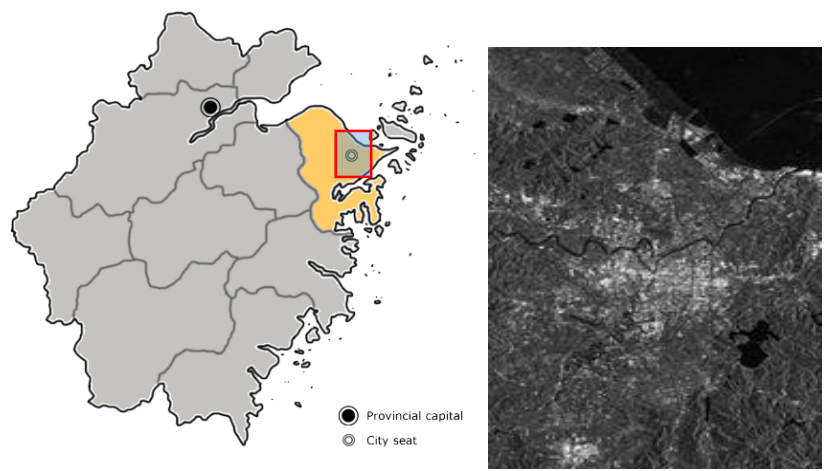

**Figure 1. The studied area**

**3.  The InSAR processing and the generation of DEM**
**A.  The result of flattened interferogram**

The original interferogram contains a lot of the smooth curved earth caused fringe that paralleled to the azimuth direction. Intensive fringes lead to difficulty in phase filtering and phase unwrapping, and tend to bring in errors. Therefore, before the subsequent processing, we first remove the fringes caused by smooth curved earth.

First we tried to remove the flat fringes by using the orbit parameters. Given the ellipsoid earth model, we calculated the two lengths from satellite to the earth model in the direction of the two lines of sights, i.e., two observations. Denote the difference of lengths as $DL = \text{LOS}_{master} - \text{LOS}_{slave}$, and the flat phase is

$$\phi_{flat} = \frac{4\pi}{\lambda} DL. \tag{1}$$

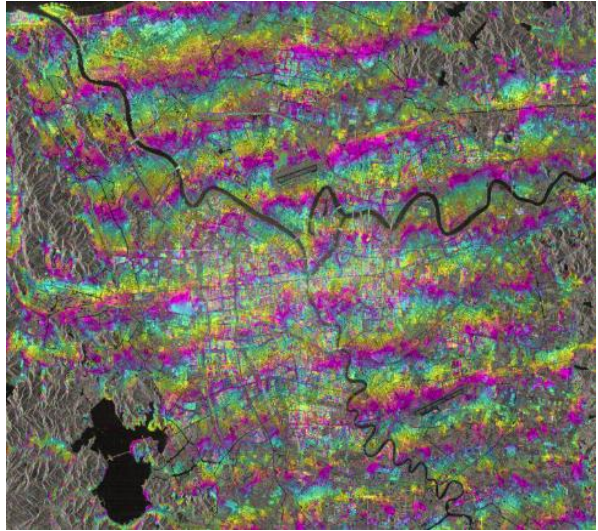

**Figure 2 Flattened interferogram by using orbit parameters**

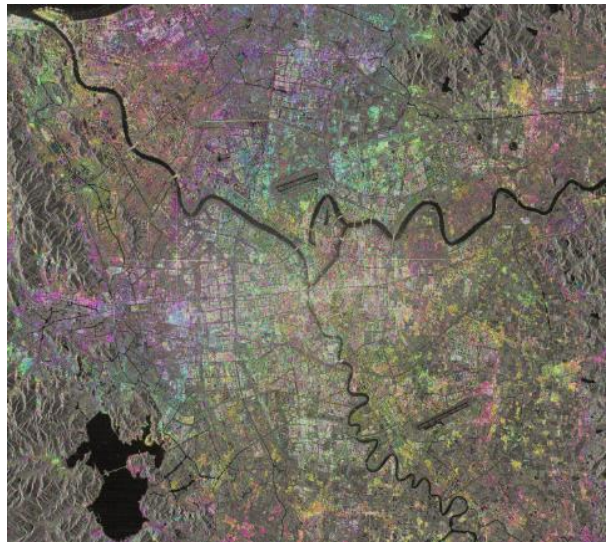

**Figure 3 Flattened interferogram by using orbit parameters by iterative baseline estimation**

However, from Figure 2, we can see that the resided flat fringes still remained. This is because the satellite orbit states we obtained were real-time and not precise enough. We noticed the relationship between the perpendicular components of the baseline and the local fringe frequency

$$f_\phi = \frac{1}{2\pi} \frac{\partial \phi}{\partial r} = -\frac{2B_\perp}{\lambda r \tan\theta} \tag{2}$$

where $\lambda$ is the wavelength, $r$ is the slant distance, $\theta$ is the local incidence angle. Using the local fringe frequency to calculate $B_\perp$ and using the $B_\perp$ to estimate the flat fringes, iteratively, and after two times of baseline estimation and fringes flatting, the interferogram changed a little. Then, we obtained the flattened interferogram. The flattened interferogram using iterative baseline estimation is shown in Figure 3.

### B. Coherence enhancement

The coherence of the original flattened interferogram is low. High coherence is the guarantee of InSAR processing. However, owing to the atmospheric propagation error, baseline, and other decorrelation factors, the coherence of the original received data cannot meet the requirements, which makes the InSAR processing difficult to carry on, even difficult to extract accurate information. Therefore, the improvement of coherence is the basic problem of InSAR processing. In this case, we put forward a kind of robust coherence enhancement interferogram filter to guarantee denoising and texture detail preservation simultaneously [3]. This method has been applied to the InSAR processing of a variety of satellite platforms, and has compared with the existing commercial software. From Figure 4(a) and (b), we can see that the mean coherence of the original interferogram map has grown from 0.42 to 0.88.

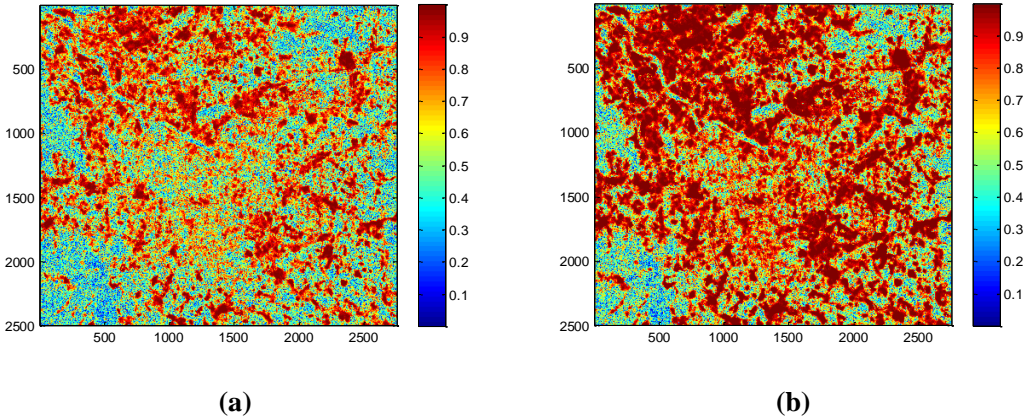

(a)          (b)

**Figure 4 The coherence maps of (a) the filtered result of the adaptive filter provided in commercial software; (b) the filtered result of the BLSURE.**

### C. Phase unwrapping

After phase filtering, most of the area's coherence is above 0.8, providing good condition for phase unwrapping. Before the inversion of the unwrapped phase, we first mask the area whose coherence is below 0.5, in order to avoid the difficulty of phase unwrapping caused by dense residues in low coherence. Here, we choose Minimum Cost Flow (MCF) method for phase unwrapping [4], however, the unwrapped results of the commercial software contain inconsistent phenomenon. Then, we use the method with Delaunay triangulation network. The unwrapped value of low coherence area can be derived according to the pixels of high quality. Through the Delaunay triangle processing, there are still some inconsistent phenomenon. To this issue, we use the phase unwrapping method based on phase-gradient-jump connections provided lately [5]. This method cannot unwrap the phase entirely in case of some complex conditions, however, it has the advantage of fast and consistent. Before the phase unwrapping by MCF, we apply the proposed method to the residues map, and create a consistent solution map. In the phase-gradient-jump connection approach, the adjacent residuals are eliminated first, and residues with different values, i.e., 1 or -1, have fixed connection. In this way, we blocked the interferogram map first, and then used the phase-gradient-jump connections method to get a consistent connection in the edge of the blocks. After that, by using MCF method with Delaunay triangulation network, we got the unwrapped phase map of each block. Finally, the entire unwrapped map can be acquired by simply jointing the blocks together.

### D. The precise baseline estimation and the inversion of heights

In section 3-A, to remove the flat phase, we estimated the baseline using several methods. That was sufficient to flatten the interferogram. Further, the inversion of height requires a more accurate estimate.

Here we choose the method of GCP based baseline estimation. In order to obtain an accurate baseline estimate by using the unwrapped phase and the ground control points (GCP), we used the equation that represents the phase change caused by height

$$\phi_{unw} = \frac{4\pi B_\perp}{\lambda r \sin\theta} P, \tag{3}$$

where $P$ is the height of GCP. Then the $B_\perp$ can be calculated as,

$$B_\perp = \frac{\lambda r \sin\theta}{4\pi P} \phi_{unw}. \tag{4}$$

The $B_\perp$ can be estimated accurately by least square fitting of the unwrapped phases and GCPs.

Here, we converted the 90 m SRTM DEM in Ningbo area to the Gaofen-3 SAR coordinates by geocoding as shown in Figure 5 and extracted 200 × 200 pixels as GCPs uniformly distributed in the phase map.

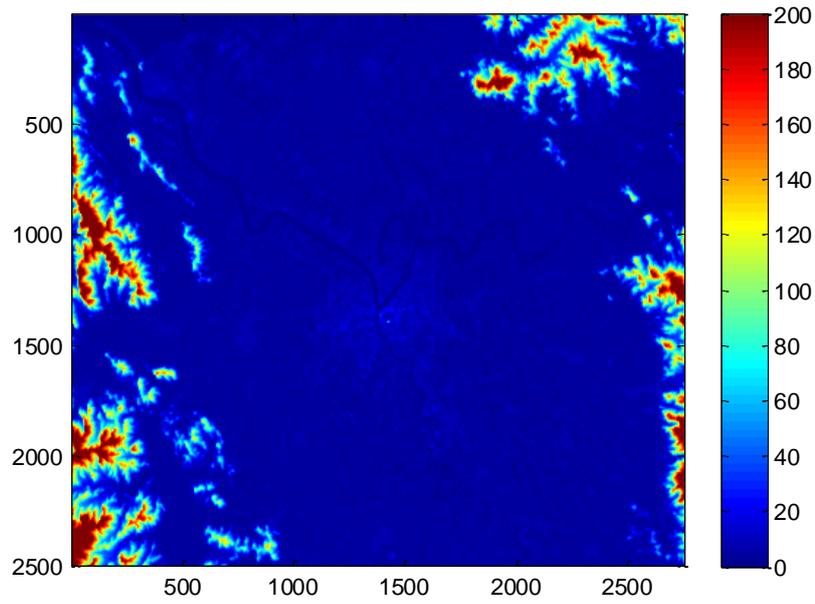

**Figure 5** SRTM DEM maps based on Gaofen-3 SAR imagery.

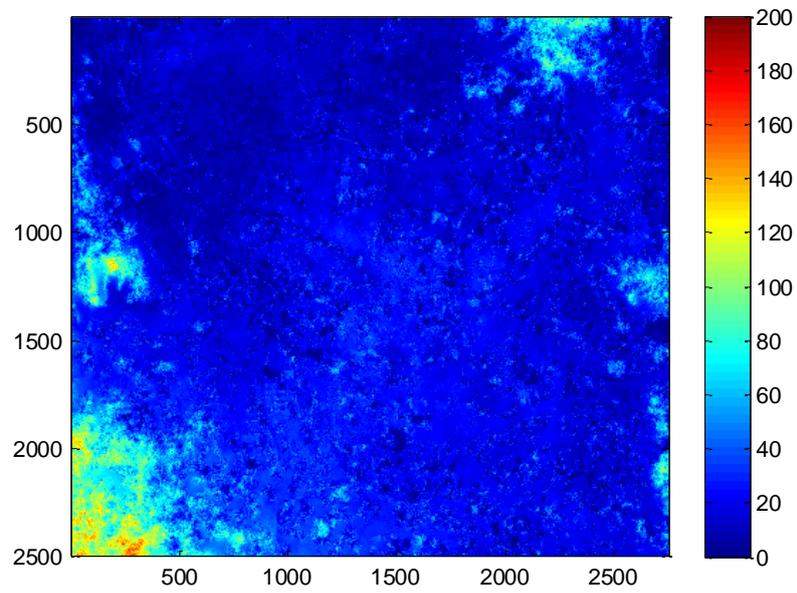

**Figure 6** DEM maps from repeat-pass Gaofen-3 SAR interferometry.

After the precise baseline estimation, the height can be obtained

$$h = \frac{\lambda r \sin\theta}{4\pi B_{\perp}} \phi_{unw},  \quad (5)$$

where $h$ is the height retrieved.

Finally, transforming the height map from radar geometry to ground range, we obtain the repeat-pass DEM as shown in Figure 6.

## 4. The evaluation of the DEM

In this section, we present the error of the elevation height in Ningbo area compared with the SRTM DEM. Table 1 shows the DEM values of Gaofen-3 and SRTM in plane area. From Table 1, we can see the RMS error is 4.5m. Table 2 shows the DEM values in mountainous region, from which we can see the RMS error is 22.3m.

Table 1. DEM Comparisons in Plane Area

| No. | SRTM DEM | Gaofen-3 DEM | Errors |
|---|---|---|---|
| 1 | 12.76 | 3.06 | -9.70 |
| 2 | 8.94 | 5.19 | -3.75 |
| 3 | 9.63 | 2.61 | -7.02 |
| 4 | 13.17 | 27.24 | 14.07 |
| 5 | 15.86 | 4.52 | -11.34 |
| 6 | -7.13 | 5.42 | 12.55 |
| 7 | 9.21 | 6.76 | -2.44 |
| 8 | 4.02 | 4.47 | 0.45 |
| 9 | 5.43 | 4.77 | -0.67 |
| 10 | 12.68 | 8.35 | -4.33 |
| 11 | 13.50 | 3.20 | -10.29 |
| 12 | 9.59 | 4.66 | -4.93 |
| 13 | 6.95 | 4.34 | -2.62 |
| 14 | 12.66 | 4.84 | -7.82 |
| 15 | -3.69 | 5.37 | 9.06 |
| 16 | 3.85 | 5.00 | 1.15 |
| 17 | 11.06 | 3.35 | -7.70 |
| 18 | 4.61 | 3.85 | -0.76 |
| 19 | 14.70 | 4.70 | -10.00 |
| 20 | 18.27 | 3.71 | -14.57 |
| 21 | 17.34 | 7.52 | -9.83 |
| 22 | 18.13 | 7.19 | -10.94 |

Table 2. DEM Comparisons in Mountainous Area

| No. | SRTM DEM | Gaofen-3 DEM | Errors |
|---|---|---|---|
| 1 | 108.41 | 72.06 | -36.34 |
| 2 | 94.18 | 92.06 | -2.12 |
| 3 | 88.81 | 103.29 | 14.47 |
| 4 | 140.93 | 125.98 | -14.94 |
| 5 | 124.69 | 143.74 | 19.04 |
| 6 | 86.53 | 117.95 | 31.42 |
| 7 | 53.96 | 54.43 | 0.47 |
| 8 | 69.50 | 46.48 | -23.03 |
| 9 | 107.89 | 63.78 | -44.11 |
| 10 | 70.94 | 23.12 | -47.82 |
| 11 | 62.50 | 5.15 | -57.36 |
| 12 | 91.48 | 59.11 | -32.37 |
| 13 | 94.43 | 118.65 | 24.22 |
| 14 | 34.55 | 5.00 | -29.55 |
| 15 | 57.70 | 5.00 | -52.70 |
| 16 | 61.95 | 18.06 | -43.89 |
| 17 | 73.92 | 41.38 | -32.55 |
| 18 | 19.83 | 27.26 | 7.43 |
| 19 | 27.37 | 5.00 | -22.37 |
| 20 | 51.38 | 5.00 | -46.38 |
| 21 | 25.62 | 5.00 | -20.62 |
| 22 | 28.91 | 11.36 | -17.55 |
| 23 | 18.76 | 6.30 | -12.46 |
| 24 | 40.29 | 24.75 | -15.55 |
| 25 | 29.48 | 6.41 | -23.07 |
| 26 | 36.38 | 5.34 | -31.04 |
| 27 | 30.14 | 6.59 | -23.56 |
| 28 | 35.28 | 5.03 | -30.25 |
| 29 | 43.53 | 3.71 | -39.82 |
| 30 | 39.45 | 5.39 | -34.06 |
| 31 | 53.35 | 84.25 | 30.90 |
| 32 | 66.26 | 37.76 | -28.50 |
| 33 | 26.13 | 3.65 | -22.48 |
| 34 | 37.63 | 5.13 | -32.50 |
| 35 | 42.80 | 4.43 | -38.36 |
| 36 | 21.33 | 5.01 | -16.32 |
| 37 | 41.24 | 12.28 | -28.96 |
| 38 | 23.45 | 3.94 | -19.51 |
| 39 | 35.56 | 4.20 | -31.36 |
| 40 | 16.05 | 4.68 | -11.36 |
| 41 | 37.53 | 13.04 | -24.49 |
| 42 | 40.99 | 6.05 | -34.94 |

## 5. Conclusion

This paper reports the first SAR interferometry experiments by using Chinese spaceborne SAR, GF-3. We analyzed the baseline estimation, coherence enhancement and phase unwrapping methods that were suitable for the data in Ningbo area. The DEM was generated after the InSAR processing, and we evaluated the elevation model by comparing with the elevation value extracted from SRTM. And the mean square error of the height is round 5 m. In summary, the InSAR processing results show that Chinese Gaofen-3 SAR satellite has the ability to achieve good elevation model in large area.

**6. Acknowledgment**

The authors thank the support from China Academy of Space Technology (CAST), China Center for Resources Satellite Data and Application (CRESDA), and National Disaster Reduction Center of China (NDRCC).